%% file: lattice_2015.tex
\documentclass{PoS}
\usepackage{booktabs}
\usepackage{amsmath}
\usepackage{bm}
\usepackage{tikz}
\usepackage{pgflibrarysnakes}
\usetikzlibrary{decorations.markings}

\newcommand{\bra}[1]{\mathinner{\langle{#1}|}}
\newcommand{\ket}[1]{\mathinner{|{#1}\rangle}}
\newcommand{\Nf}{N_{\mathrm f}}
\newcommand{\Oa}{\mathrm{O}(a)}
\newcommand{\Ep}{E_{\bm p}}

\newcommand{\Eval}[1]{\big\langle #1 \big\rangle}

\title{Nucleon electromagnetic form factors and axial charge from CLS $N_\mathrm{f}=2+1$ ensembles}

\ShortTitle{Nucleon EM form factors and $g_A$ from CLS $\Nf=2+1$ ensembles}

\author{Dalibor Djukanovic, \speaker{Tim Harris}, Parikshit Junnarkar\\
        Helmholtz Institute Mainz\\
        Johann-Joachim-Becher-Weg 36\\
        55128 Mainz

        E-mail: \email{harris@kph.uni-mainz.de}
    }

\author{Georg von Hippel, Harvey B. Meyer, Hartmut Wittig\\
       PRISMA Cluster of Excellence and Institute for Nuclear Physics\\
       University of Mainz\\
       Johann-Joachim-Becher-Weg 45\\
       55128 Mainz
   }

\abstract{We present preliminary results on the electromagnetic form factors and axial charge of the nucleon from ensembles generated by the CLS effort with $N_\mathrm{f}=2+1$ flavours of non-perturbatively O($a$)-improved Wilson fermions and open temporal boundary conditions.
        Systematic effects due to excited-state contamination are accounted for using both two-state fits and the method of summed operator insertions.
        This exploratory analysis demonstrates the viability of obtaining precision baryon observables with $\Nf=2+1$ flavours of Wilson fermions on fine lattices, aiming towards controlled chiral and continuum limits in the future.
    }

\FullConference{The 33rd International Symposium on Lattice Field Theory\\
        14 -18 July 2015\\
        Kobe International Conference Center, Kobe, Japan*
    }

\begin{document}

\section{Introduction}
The form factors of the nucleon and its coupling to external currents are important quantities which contribute to our understanding of hadron structure through, for example, charges and their distributions.
Reliable non-perturbative determinations of these quantities from lattice QCD are valuable to address, for example, the proton radius puzzle~\cite{Antognini25012013}, and have comprised part of the programme of many groups~\cite{Shanahan:2014cga,Pleiter:2011gw,Abdel-Rehim:2015jna,Green:2014xba}.
Nevertheless, lattice studies have failed to provide agreement with several key experimental observables, like the axial charge $g_A=1.2723(23)g_V$~\cite{Agashe:2014kda}, possibly owing to systematic uncertainties such as finite-size, discretization and excited-state effects, which are difficult to control.
In these proceedings we report on early progress on the nucleon Sachs electromagnetic form factors,
\begin{align}
    G_E(Q^2) &= F_1(Q^2) - \frac{Q^2}{4m_N^2}F_2(Q^2), \\
    G_M(Q^2) &= F_1(Q^2) + F_2(Q^2),
\end{align}
and axial charge, $g_A=G_A(0)$, which can be accessed by computing the matrix elements of the vector $V_\mu(x)=\bar{\psi}(x)\gamma_\mu\psi(x)$ and axial $A_\mu(x)= \bar{\psi}(x)\gamma_5\gamma_\mu\psi(x)$ (isovector) currents, with a nucleon with momentum $\bm p$ and spin $s$,
\begin{align}
    \bra{N(\bm{p'},s')}V_\mu(0)\ket{N(\bm{p},s)}
        &= \bar{u}(\bm{p'},s')\left[ \gamma_\mu F_1(Q^2) + i\frac{\sigma_{\mu\nu}q^\nu}{2m_N}F_2(Q^2)\right] u(\bm{p},s),\\
    \bra{N(\bm{p'},s')}A_\mu(0)\ket{N(\bm p,s)}
        &= \bar{u}(\bm{p'},s')\left[ \gamma_5\gamma_\mu G_A(Q^2) + \frac{\gamma_5q_\mu}{2m_N}G_P(Q^2)\right] u(\bm p,s),
\end{align}
where $Q^2 =-q^2= -(p'-p)^2$ and $m_N$ is the nucleon mass.
In this work, the matrix elements are calculated using ensembles with $\Nf=2+1$ flavours of Wilson fermions, developing upon earlier $\Nf=2$ determinations by the Mainz group~\cite{Capitani:2012gj,Capitani:2015sba}.

\section{Lattice set-up}
The ensembles used in this work, listed in table~\ref{tab:cls}, have been generated with $\Nf=2+1$ flavours of non-perturbatively O($a$)-improved Wilson fermion~\cite{Bulava:2013cta} by the CLS effort~\cite{Bruno:2014jqa}.
These ensembles utilize open boundary conditions in the time direction in order to sample correctly from all topological charge sectors at fine lattice spacings~\cite{Luscher:2011kk}.
Our initial effort has focussed on ensembles at the coarsest lattice spacing, $a\approx0.086$fm, with three values of the pseudoscalar mass, $m_\pi=350$, 280 and 220 MeV, called H102, H105 and C101, respectively.
\begin{table}
    \centering
    \begin{tabular}{*{11}{c}}
        \toprule
        name    &   chains  & $\beta$ & $a$ (fm)  &   $T/a$ & $L/a$ &   $m_{\pi} L$
            &   $m_{\pi}$ (MeV) & $N_{\mathrm{cfg}}$        & $N_\mathrm{src}$        & $N_{t_s}$ \\
        \midrule
        H102    & H102r001&   3.4     & 0.086    & 96    & 32      &   5.8         &   350    &  997  & 4  & 3   \\
                & H102r002&   ''      &  ''      & ''    & ''      &   ''          &   ''     & 1000  & '' & ''  \\
        H105    & H105r002&   ''      &  ''      & ''    & ''      &   4.9         &   280    & 1000  & '' & ''  \\
                & H105r005&   ''      &  ''      & ''    & ''      &   ''          &   ''     &  837  & '' & ''  \\
        C101    & C101r015&   ''      &  ''      & ''    & 48      &   4.7         &   220    &  575  & 8  & ''  \\
        \midrule  
        N200    & N200r000&   3.55    & 0.064    & 128   & ''      &   4.4         &   280    &  800  & 4  & 6 \\
        \bottomrule
    \end{tabular}
    \caption{Details of the CLS $\Nf = 2+1$ ensembles and measurements used in this work.}
    \label{tab:cls}
\end{table}
\subsection{Calculation of the matrix elements}
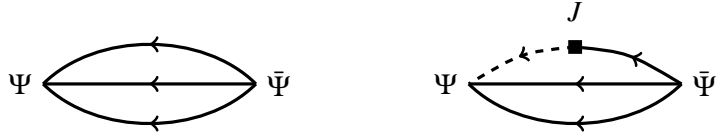
\begin{figure}[h]
    \centering
    \input{threept}
    \caption{Schematic contractions for the connected contributions to the nucleon two- (left) and three-point (right) functions.}
    \label{fig:threept}
\end{figure}
In order to optimize the overlap of the interpolating operator with the ground state, we employ Wuppertal smearing with APE smeared links~\cite{Gusken:1989qx,Albanese:1987ds}, with parameters chosen to optimize the effective mass plateau of the nucleon at short distances.
The two- and three-point functions,
\begin{align}
    C_{2}(t;\bm q)
    &= \Gamma_{\alpha\beta}\sum_{\bm x} e^{-i\bm q\bm x}\Eval{\Psi_\beta(\bm x, t)\bar{\Psi}_\alpha(0)}, \\
    C_{3,J}(t,t_s;\bm q)
    &= \Gamma_{\alpha\beta}\sum_{\bm x,\bm y} e^{- i\bm q\bm y}\Eval{\Psi_\beta(\bm x, t_s)J(\bm y, t)\bar{\Psi}_\alpha(0)},
    \label{eq:nucleon_corr}
\end{align}
with polarization matrix $\Gamma=(1+\gamma_0)(1+i\gamma_5\gamma_3)$, are illustrated in figure~\ref{fig:threept}.
The three-point function is calculated using the sequential source method, where the dashed propagator in figure~\ref{fig:threept} (right) is computed independently of the current, $J$, and the final-state nucleon is always at rest.
In the asymptotic regime dominated by the ground state, the ratios
\begin{align}
    R_J(t,t_s;Q^2) = \frac{C_{3,J}(t,t_s;\bm q)}{C_2(t_s;\bm q)}\sqrt{\frac{C_2(t_s-t;-\bm q)C_2(t,\bm 0)C_2(t_s;\bm 0)}{C_2(t_s-t;\bm 0)C_2(t;-\bm q)C_2(t_s;-\bm q)}},
\end{align}
have been demonstrated~\cite{Alexandrou:2008rp} to have good overlap properties to access the form factors and axial charge from the (Euclidean) isovector currents
\begin{align}
    \mathrm{Re} R_{V^E_0}(t, t_s; Q^2) &= \sqrt{\frac{m_N + \Ep}{\Ep}} G_E(Q^2), \\
    \mathrm{Re} R_{V^E_i}(t, t_s; Q^2) &= \frac{\epsilon^{ij}q^j}{\sqrt{2\Ep(\Ep+m_N)}}G_M(Q^2), \quad i,j\in\{1,2\},\\
    \mathrm{Im} R_{A^E_{3}}(t, t_s; 0) &= g_A,
\end{align}
where $\Ep^2 = \bm p^2 + m_N^2$ and $\epsilon^{ji}=-\epsilon^{ij}$, $\epsilon^{12}=+1$.
Note that we use the exactly-conserved (point-split) discretization of the vector current.

The boundary conditions break translation invariance in the time direction, and correlation functions receive contributions which decay at least with energy of twice the pseudoscalar mass times the distance from the operator to the nearest boundary, which is the lowest energy with vacuum quantum numbers.
For example, in the two-point function with source located on at a distance $t_\mathrm{src}/a$ from the boundary,
\begin{align}
    C_2(t,\bm 0) 
        = \lvert\bra{\emptyset}\Psi\ket{N}\rvert^2 e^{-m_N t}
        + \bra{\emptyset}\Psi\ket{N}\bra{N}\bar \Psi\ket{2\pi} e^{-2m_\pi t_\mathrm{src}}
            + \ldots.
    \label{eq:bnd}
\end{align}
To minimize such effects, we have adopted a conservative approach to fix the source location such that for the largest source-sink separation both operators are maximally distant from the boundaries.
For example, with $T/a=96$ and maximum source-sink separation of $t_s/a=16$, we choose the source temporal coordinate $t_\mathrm{src}/a=40$.
In total four sources are used per configuration, placed at one corner and the face-centres of the cubic spatial volume.
Additional sources for the C101 ensemble are placed in the cubic body- and edge-centres.

\subsection{Accounting for excited states}
To eliminate residual excited-state effects, we have used three source-sink separations in this work with $t_s/a\in\{12,14,16\}$, corresponding to $t_s\approx1.0$--1.4fm.
Following~\cite{Capitani:2015sba} we employ two methods to control excited-state effects:
\begin{enumerate}
    \item \emph{two-state method}: the form-factor $\hat{G}_X(Q^2)$ is estimated by modelling the contribution of the leading excited state,
        \begin{align}
            G^\mathrm{eff}_X(t,t_s;Q^2)
            = \hat{G}_X(Q^2) +  a_X(Q^2)e^{-\Delta t} + b_X(Q^2)e^{-\Delta'(t_s-t)}.
            \label{eq:two_state}
        \end{align}
        In this case, the energy gap is fixed to the lowest non-interacting level, $\Delta=m_\pi$ or $2m_\pi$ and $\Delta'=2m_\pi$ and leaving the transition matrix elements, $a_X$, $b_X$ as free parameters.
        In the case of the axial charge, the couplings are subject to the constraint $a_X=b_X$.
        Simultaneous fits are performed to all source-sink separations.

        Recent work by members of our group has investigated the approximation of using the non-interacting levels.
        By using experimental scattering data in the L\"uscher finite-volume formalism, it appears that the interacting $N\pi$ levels in a finite volume are close to their non-interacting counterparts~\cite{Hansen_note}.
        This gives us some confidence in the plausibility of such an ansatz.

        However, recent predictions from leading-order chiral perturbation~\cite{Bar:2015zha} theory suggest that the coupling to the first excited-state should be significantly smaller than what is typically observed when fixing the gap to the first non-interacting level.
        This leading-order prediction would induce an overestimation of the axial charge if this leading excited-state contamination were not accounted for.
        Typically, an underestimation is reported from lattice data.

    \item \emph{summation method}: by summing over the position of the temporal coordinate of the operator insertion, an estimator with excited-state effects $\mathrm{O}(e^{-\Delta t_s})$ can be obtained from the slope of the function as
        \begin{align}
            \sum_{t=0}^{t_s}G_X^{\mathrm{eff}}(t,t_s; Q^2)\stackrel{t_s\gg 0}{\rightarrow}
                 K_X(Q^2)+t_s\hat G_X(Q^2) + \ldots,
            \label{eq:summation}
        \end{align}
        where $K_X(Q^2)$ denote (generally divergent) constants and omitted terms are exponentially suppressed.
        However, this can result in a trade-off between the statistical and systematic error as the statistical error on the sum grows with $t_s$ and the slope may not be accurately determined.
\end{enumerate}

\section{Electromagnetic form factors}
\begin{figure}[t]
    \centering
    \includegraphics[scale=0.5]{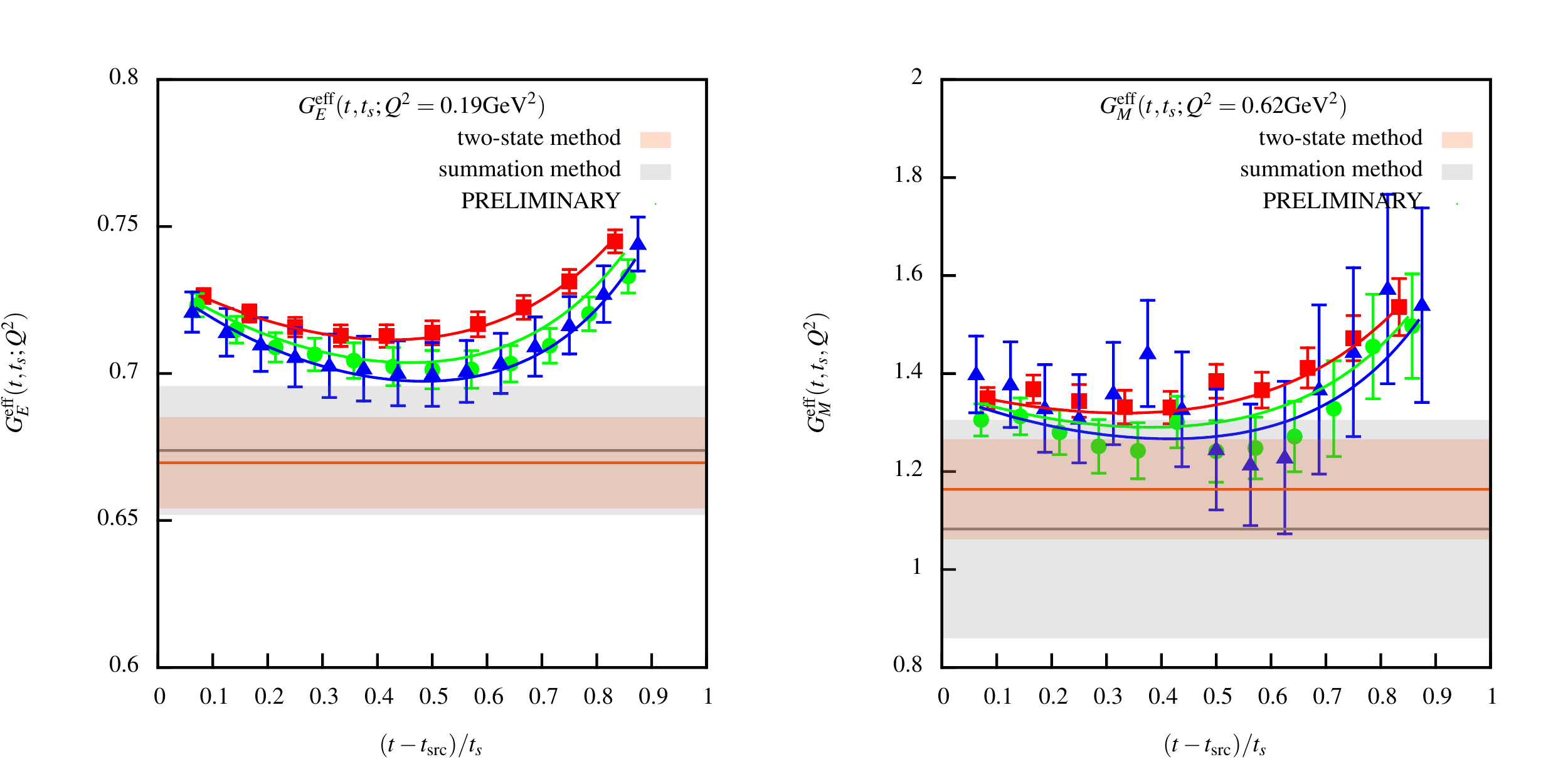}
    \caption{Effective electric (left) and magnetic (right) form factors at $Q^2=0.19\mathrm{ GeV}^2$ and $Q^2=0.62\mathrm{ GeV}^2$ on the H102 ensemble central values and errors from the two-state (orange band) and summation (grey band) methods.}
    \label{fig:effff_H102}
\end{figure}

Examples of the effective form-factors, $G_E(t,t_s;Q^2)$ and $G_M(t,t_s;Q^2)$, for the H102 ensembles with pseudoscalar mass $m_\pi=350$ MeV are shown in figure~\ref{fig:effff_H102} in the left and right panels respectively.
The red squares, green circles and blue triangles are the data for source-sink separations $t_s/a=12,14,16$ respectively.
The characteristic curvature attributed to non-perfect overlap with the nucleon ground-state is clearly seen.
In particular, the slower decay of these effects at the source than at the sink is consistent with our expectations from the coupling to the different excited states due to the kinematical set-up.

The central value and error from the two-state method are shown with the orange line and band.
The band is observed not to be compatible with any of the data themselves, which suggests the significance of the excited-state contribution.
Nevertheless, the model describes the data well as can be observed from the solid red, green and blue lines.
The grey solid line and band correspond to the central value and error from the summation method.
Both methods provide good descriptions of the data with uncorrelated $\chi^2/\mathrm{dof}<1$.
Good agreement is observed between both methods however the estimation from the summation method is not as precise as the two-state method.

\begin{figure}[t]
    \centering
    \includegraphics[scale=0.5]{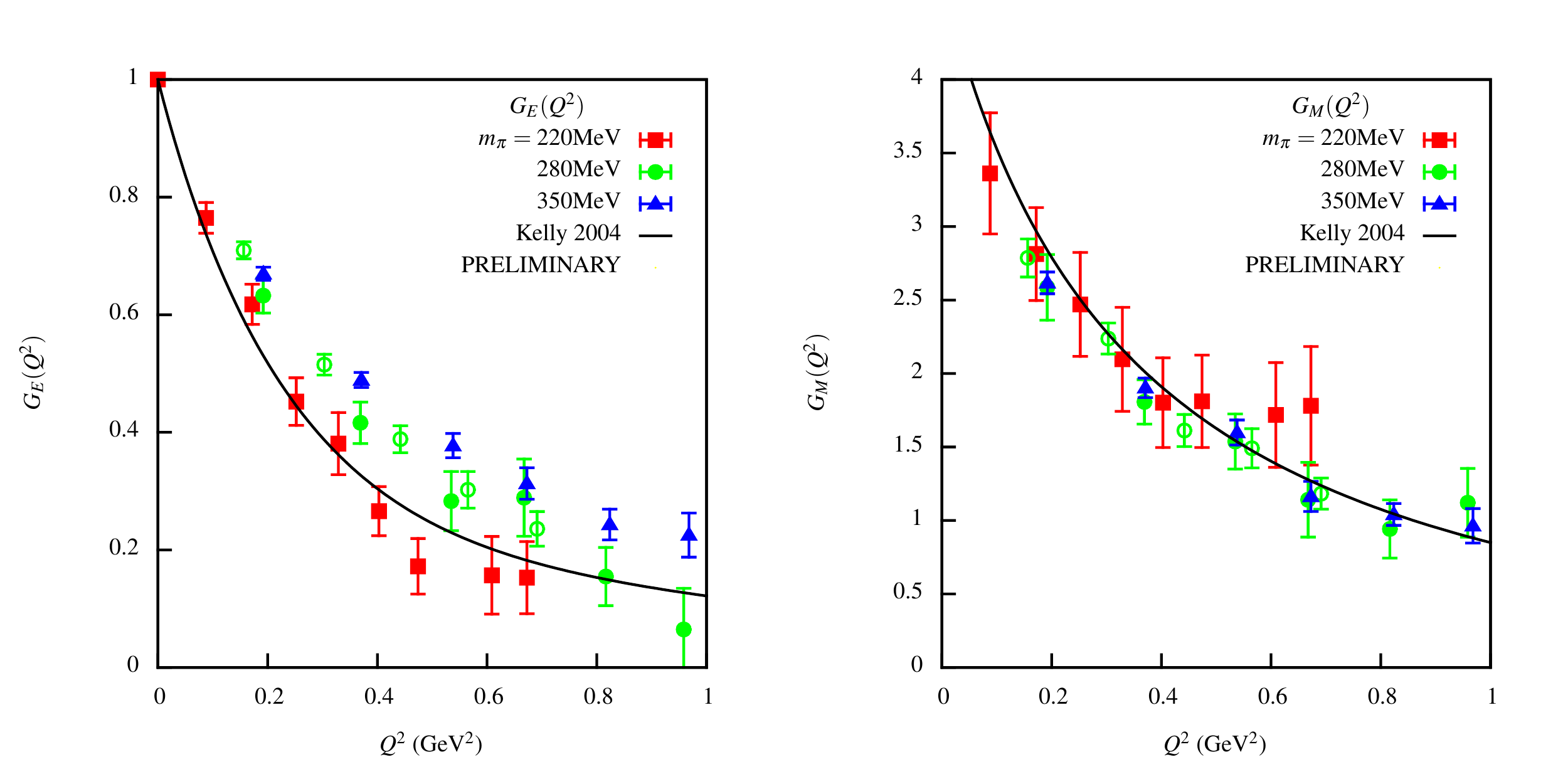}
    \caption{Preliminary electric (left) and magnetic (right) form factors estimated using the two-state method with the phenomenological curve parameterized by Kelly.}
    \label{fig:ff_beta3p43p55}
\end{figure}
In order to compare the dependence on the pseudoscalar mass, we compile the results from the two-state method from the different $\beta=3.4$ ensembles (full symbols) in figure~\ref{fig:ff_beta3p43p55}.
The solid black line depicts the parameterization of the experimental data by Kelly~\cite{Kelly:2004hm}, which at the current precision cannot be distinguished from more recent parameterizations~\cite{Bernauer:2013tpr}.
Strong dependence on the pseudoscalar mass is observed in the electric form factor (figure~\ref{fig:ff_beta3p43p55}, left), which may lead to an undershooting of the experimental curve after an extrapolation to the physical pion mass.
In contrast, the magnetic form factor (figure~\ref{fig:ff_beta3p43p55}, right) shows little dependence on the pseudoscalar mass.
The preliminary data for the N200 ensemble with $m_\pi=280$ MeV at a finer lattice spacing are displayed in figure~\ref{fig:ff_beta3p43p55} (open symbols), which are in good agreement with the $\beta=3.4$ data at a similar value of the $m_\pi$, meaning lattice artifacts cannot be resolved with the current precision.

\begin{figure}[t]
    \centering
    \includegraphics[scale=0.5]{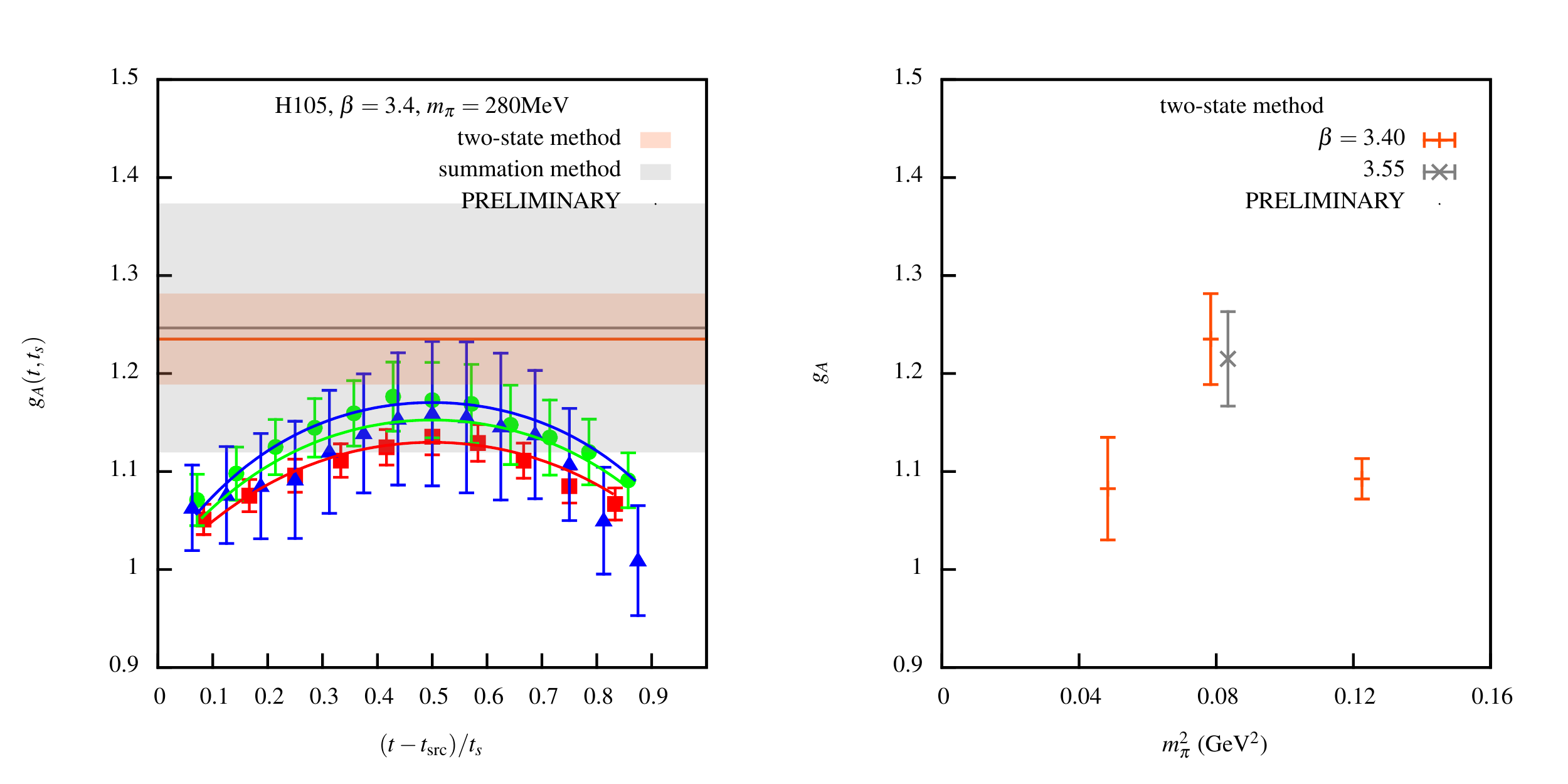}
    \caption{Nucleon axial charge on H105 ensemble (left) and dependence on the pseudoscalar mass.}
    \label{fig:gA}
\end{figure}

\section{Axial charge}
Preliminary results for the axial charge are shown in figure~\ref{fig:gA}.
The effective axial charge is shown in the left-hand panel where the red squares, green circles and blue triangles are the data for source-sink separations $t_s/a=12,14,16$ respectively.
Again, the orange and grey bands depict the axial charge estimated using the two-state method and summation method.
Similar to the case of the electromagnetic form factors, the central value of the axial charge lies outside the error of the data.
The error on the summation method is significantly larger than that of the two-state method.
For comparison of the different ensembles, in the right-hand panel of figure~\ref{fig:gA} we show the axial charge as a function of the pseudoscalar mass, although no discernible trend is visible.

\section{Summary}
In this preliminary report we have investigated the nucleon electromagnetic form factors and axial charge on $\Nf=2+1$ flavours of $\Oa$-improved Wilson fermions, extending our $\Nf=2$ determinations.
The open boundary conditions pose no particular extra challenge.
In order to improve the precision of these observables we are currently exploring the use of the truncated-solver method in order to perform more efficient measurements~\cite{Shintani:2015aea,Bali:2015qya}.
Additionally, more sophisticated analyses of the excited-state effects are underway, as well as the $\Oa$-improvement of the currents themselves.
In future, we aim to cover a broader parameter space in the pseudoscalar mass and lattice spacing before parameterizing the data at the physical point.
{\small
\paragraph{Acknowledgments}
We are grateful to our colleagues within the CLS initiative for sharing ensembles.
The programs were written using QDP++~\cite{Edwards:2004sx} with the deflated SAP+GCR solver from openQCD~\cite{openQCD}.
The correlation functions were computed on the ``Clover'' platform at the Helmholtz Institute Mainz.
}

\bibliography{lattice_2015}

\end{document}

%% file: threept.tex
\begin{tikzpicture}[scale=0.7]
    \begin{scope}[very thick,decoration={
        markings,
        mark=at position 0.5 with {\arrow{>}}}
        ] 
        \begin{scope}[shift = {(4,0)}]
        
        \draw[fill] (-2.1,0.6) rectangle (-1.9,0.8);
        \draw[very thick, postaction={decorate}] (0,0) .. controls (-1,0.6) .. (-2,0.7) node[above=0.5em]{$J$};
        \draw[very thick, dashed, postaction={decorate}] (-2,0.7) .. controls (-3,0.6) .. (-4,0);
        \draw[very thick, postaction={decorate}] (0,0) .. controls (-1,-1) and (-3,-1) .. (-4,0);
        \draw[very thick, postaction={decorate}] (0,0) node[right]{$\bar \Psi$} -- (-4,0) node[left]{${\Psi}$};
    \end{scope}
        \begin{scope}[shift = {(-4,0)}]
            \draw[very thick, postaction={decorate}] (0,0) .. controls (-1,+1) and (-3,+1) .. (-4,0);
            \draw[very thick, postaction={decorate}] (0,0) .. controls (-1,-1) and (-3,-1) .. (-4,0);
            \draw[very thick, postaction={decorate}] (0,0) node[right]{$\bar{\Psi}$} -- (-4,0) node[left]{$\Psi$};
        \end{scope}
    \end{scope}
\end{tikzpicture}